\documentclass[11pt]{article} 

\usepackage[utf8]{inputenc} 
\usepackage{setspace}
\usepackage{color}
\usepackage{url}

\usepackage{geometry} 
\usepackage{graphicx} 

\usepackage{booktabs} 
\usepackage{array} 
\usepackage{paralist} 
\usepackage{verbatim} 
\usepackage{subfig} 
\usepackage{authblk}

\usepackage{fancyhdr} 
\usepackage{sectsty}
\allsectionsfont{\sffamily\mdseries\upshape} 

\usepackage[nottoc,notlof,notlot]{tocbibind} 
\usepackage[titles,subfigure]{tocloft} 

\newcommand\green[1]{\textcolor{green}{ #1}}

\title{Disruption of transfer entropy and inter-hemispheric brain functional connectivity  in patients with disorder of consciousness}
\author[1,2,5]{Ver\'onica M\"aki-Marttunen}
\author[3,5]{Ibai Diez}
\author[3,4]{Jesus M. Cortes}
\author[2]{Dante R. Chialvo}
\author[1,2,*]{Mirta Villarreal}
\affil[1]{Department of Cognitive Neuroscience. FLENI, Buenos Aires, Argentina}
\affil[2]{CONICET. Buenos Aires, Argentina}
\affil[3]{Biocruces Health Research Institute. Hospital Universitario de Cruces. Barakaldo, Spain}
\affil[4]{Ikerbasque, The Basque Foundation for Science.  Bilbao,  Spain}
\affil[5]{Equal contribution}
\affil[*]{Corresponding author: mvillarreal@fleni.org.ar }

\begin{document}
\maketitle
\begin{abstract}
Severe traumatic brain injury can lead to disorders of consciousness (DOC) characterized by deficit    in conscious awareness and cognitive impairment including coma, vegetative state, minimally consciousness, and lock-in syndrome. 
Of crucial importance is to find objective markers that can account for the large-scale disturbances of brain function to help the diagnosis and prognosis of DOC  patients and eventually the prediction  of  the coma outcome. 
Following recent studies suggesting that the functional organization of brain networks can be altered in comatose patients,  this work analyzes  brain functional connectivity (FC)  networks obtained from resting-state functional magnetic resonance imaging (rs-fMRI).  Two approaches are used to estimate the FC: the Partial Correlation (PC) and the Transfer Entropy (TE). 
Both the PC and the TE   show significant statistical differences between the group of patients and   control subjects; in brief,   the inter-hemispheric PC and the   intra-hemispheric TE   account  for such    differences. 
Overall, these  results  suggest  two possible rs-fMRI markers useful to design new strategies for the management and neuropsychological rehabilitation of DOC patients. 
\end{abstract}

\clearpage

\section*{Introduction}

Recent studies have shown that brain networks obtained from functional Magnetic Resonance Imaging (fMRI) recordings  are altered in patients with severe disorder of consciousness (DOC)\cite{boveroux2010,noirhomme2010,heine2012,diperri2013}. DOC can result from severe brain injury and is characterized by an absence of awareness of the self and the environment, either with preserved or disrupted sleep-awake cycle.  DOC encompasses a wide spectrum of clinical conditions with different levels in the content of conscious awareness, ranging from the coma state (CS, patients who have a disrupted sleep-awake cycle and don't wake up), vegetative state (VS, who preserve sleep-awake cycle but are unaware of themselves and the environment), minimally consciousness state (MCS, patients who are unable to reliably communicate but show reproducible albeit fluctuating behavioral evidence of awareness), to lock-in syndrome (LI, patients who are fully conscious but are completely paralyzed except for small movements of the eyes or eyelids). For  the prognosis of these patients, the clinical practice scores this graduation in DOC response   by the Glasgow Coma Scale (GCS) \cite{teasdale1974}, or as we will use in this paper, by an alternative  scale such as the JFK   Coma Recovery Scale-Revised (CSR-R)  \cite{giacino2004}. This scale encodes  the neurological and behavioural state of the DOC patient providing  a number    ranging from 0 to 23, 0 for the deepest coma state, 23 for the fully recovered one. 
Despite the existence of such  scales, there is a  need  for  more reliable methods that based on brain neuroimaging  can provide     better characterization of  the  large-scale disturbances of brain function in DOC. Ultimately these approaches should help in understanding and eventually predicting coma outcome. 

The resting state functional Magnetic Resonance Imaging (rs-fMRI)  accounts  for the spontaneous  brain activity  occurring in the  high-amplitude ultra-slow (0.1 Hz) fluctuations in the Blood-Oxygen-Level-Dependent (BOLD) signal, defining networks   of correlated spontaneous activity  of  brain Functional Connectivity (FC) \cite{raichle2001,beckmann2005}. The interaction between these distributed networks as well as subcortical modules  is considered critical for conscious processing, and has been shown to be disrupted in DOC state \cite{tononi2004,rosanova2012, cauda2009}.  Furthermore, the rs-fMRI paradigm is a very suitable strategy for DOC patients, since they  are not able to efficiently perform  specific tasks.
The present study addresses the question of whether the FC obtained from the rs-fMRI is altered at different brain regions as a consequence of consciousness disturbances. To this end, we investigate the FC obtained by two different measures: the Partial Correlation (PC) and the Transfer Entropy (TE), in two different groups: healthy adults and DOC patients. 

Information theory offers an arsenal of  different measures, complementing the linear correlation estimations of FC. These information tools are typically built as extensions of the Shannon Entropy, quantify the interactions between variables by measuring the information which is shared or transferred between  them \cite{jaynes1957,cover2006}.   In the last decade, the transfer entropy (TE) method is growing in popularity as it  can  account for directed interactions between time-series variables \cite{schreiber2000}.  When applied to neuroimaging time-series, TE is a data-driven measure that assesses the functional connectivity between brain areas even for  non-linear interactions. Unlike the correlations, TE reveals directionality in the interactions,  allowing for determining a \textit{directed} FC  between areas. 

We hypothesize that FC would be reduced in DOC patients since consciousness implies functional integration \cite{tononi2004}. We anticipate that PC and TE would show different behaviors in patients with increasing level of consciousness, provided that they can be related to different mechanisms of information processing in the brain.

The paper is organized as follow: in Material and Methods, we give details on the the data acquisition and preprocessing and   define the two measures PC and TE to compute FC patterns. The next section is dedicated to present the results of the analysis.  The paper closes with a discussion on some consequences of  the alteration of the FC patterns in DOC patients.

\section*{Material \& Methods}
\subsection*{Subjects}

Seventeen healthy subjects (\textbf{Group 1})  aged 25 $\pm$ 5 year old (8 men, 9 women), with no history of neurological or psychiatric problems, participated in this study as a control group. The Edinburgh Handedness Inventory was used to assess handedness \cite{oldfield1971}, resulting in  thirteen subjects  right-handed and four left-handed. 
Eleven DOC patients  (\textbf{Group 2}) were scanned (age range, 17- 44 years; 6 men, 5 women). Data from two patients were subsequently excluded because of unacceptable degrees of head and body movements. The coma severity for each patient was clinically assessed using the Revised Coma Recovery Scale (CRS-R, \cite{giacino2004}): scores range from 0 (meaning deep coma state) to 23 (full recovery). The patients were scanned the first time between 2 to 6 months after major acute brain injury, and a second time between 3 to 6 months after the first scan (Table \ref{Table1}). For better comparison, group 2 was subdivided into 2 subgroups: \textbf{Group 2a} (n = 12) is composed by all  scans of DOC patients who had a corresponding CRS-R scale. \textbf{Group 2b} (n = 4) includes the second scans of the four patients who recovered consciousness before the second session (marked with asterisks in Table \ref{Table1}).  
The study protocol was approved by the Institutional Review Board of the Institute of Neurological Research FLENI. Informed consent was directly obtained from healthy participants and from the next kin of each of the  patients.

\begin{table}
\begin{center}
\footnotesize 
\caption{\label{Table1} Clinical characteristics of DOC patients. VS: Vegetative State; MCS:  Minimally Consciousness State;   C: Conscious; EMCS: Emergence from MCS (an intermediate state between MCS and C). }
\begin{tabular}[t]{|l|l|c|c|c|c|}
\hline
Patient & Age & Time between accident& Clinical assessment & Time between first and & Clinical assessment \\
code &  & and first scan (months) & at first scan & second scan (months) & at second scan\\
\hline
P1 & 34 & 2 & VS & 5 & VS \\
\hline
P2$^{*}$ & 18 & 4 & MCS & 4 & C \\
\hline
P3$^{*}$ & 44 & 2 & MCS & 3 & C \\
\hline
P4 & 17 & 6 & VS & 6 & MCS \\
\hline
P5 & 26 & 4 & VS & 3 & MCS \\
\hline
P6$^{*}$ & 26 & 4 & EMCS & 4 & C \\
\hline
P7 & 29 & 4 & MCS & 3 & MCS \\
\hline
P8 & 41 & 2 & VS & 6 & VS \\
\hline
P9$^{*}$ & 34 & 5 & VS & 5 & C \\
\hline
\end{tabular}
\end{center}
\end{table}

\subsection*{MRI data acquisition and preprocessing}
The fMRI measurements were carried out on a 3T Signa HDxt GE scanner using an 8 channel head coil. Change in blood-oxygenation-level-dependent (BOLD) T2* signal was measured using an interleaved gradient-echo EPI sequence.  Thirty contiguous slices were obtained in the AC-PC plane with the following parameters: 2 sec repetition time (TR), flip angle: 90$^{\circ}$, 24 cm field of view, 64 x 64 pixel matrix, and 3.75 x 3.75 x 4.0 mm voxel dimensions. During the experimental session subjects lied quietly for a period of 7 minutes. 220 whole brain volumes were obtained per scan session, including 5 dummy scans to allow for T1 saturation effects that were discarded from the analysis. High resolution T1-weighted 3D fast SPGR-IR   were also acquired (TR= 6.604 ms, TE = 2.796 ms, TI = 450; parallel imaging (ASSET) acceleration factor=2; acquisition matrix size=256x256; FOV=24 cm; slice thickness=1.2 mm; 120 contiguous sections). The image data was analyzed using SPM8 (Wellcome Department of Cognitive Neurology, London, UK) implemented in MATLAB  (MathWorks Inc., Natick, MA). The functional images were subjected to temporal alignment and volumes were corrected for movement using a six-parameter automated algorithm. The realigned volumes were spatially normalized to fit to the template created using the Montreal Neurological Institute reference brain based on Talairach and Tournoux's sterotaxic coordinate system \cite{ashburner1999}. The spatially normalized volumes consisting of 4 x 4 x 4 mm$^{3}$ voxels were smoothed with a 8-mm FWHM isotropic Gaussian kernel. Additionally, a linear trend removal and band pass filtering between 0.01 and 0.08 Hz was applied on the data.

 \subsection*{Brain parcellation and Regions of Interest}
\begin{figure}[h]
\begin{center}
\includegraphics[width=.7\textwidth]{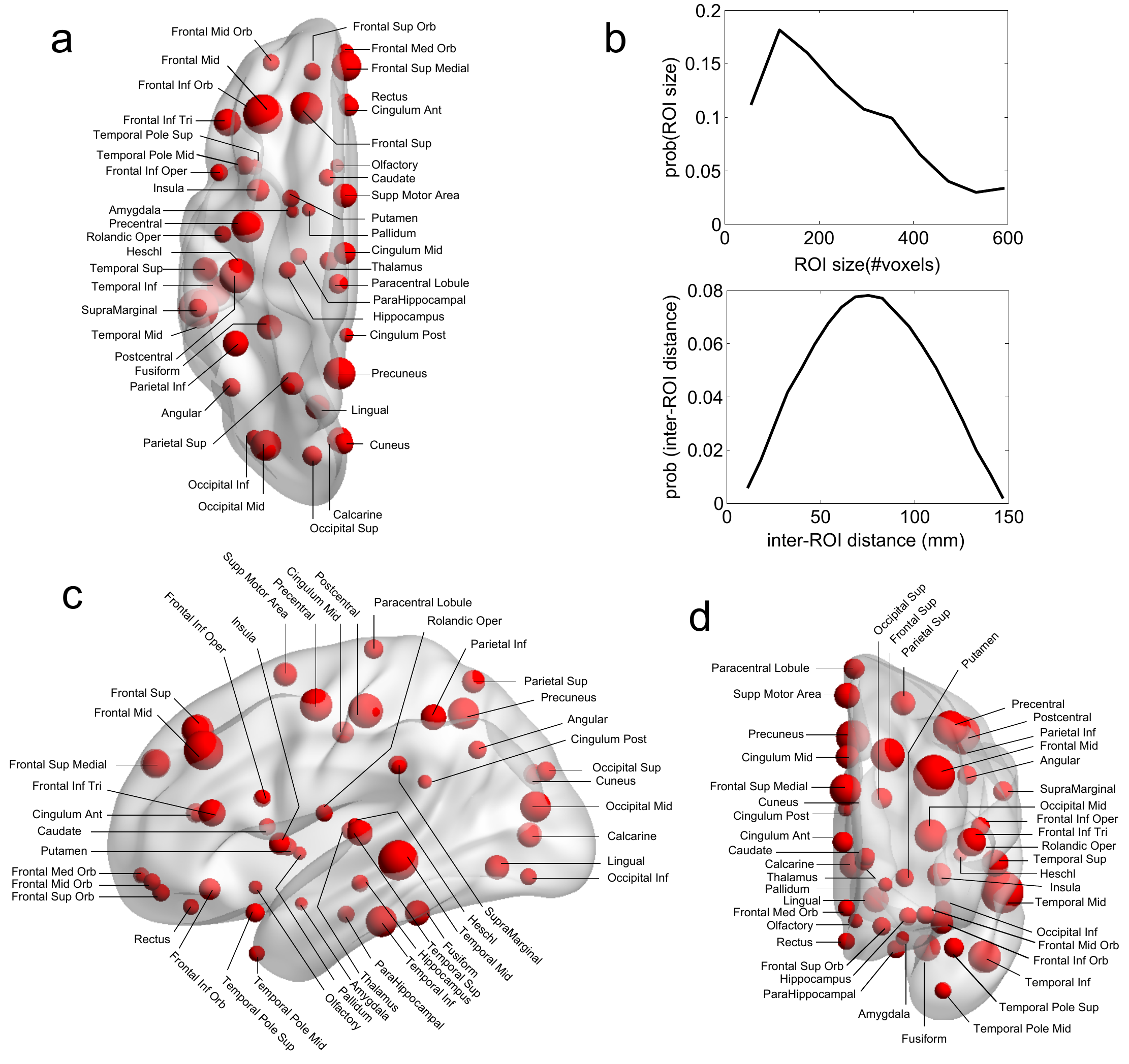}
\end{center}
\caption{{\bf  Anatomical Brain parcellation  and Regions of Interest (ROI).} \textbf{a}: Axial,  \textbf{c}: Saggital , \textbf{d}: Coronal views. Specific ROI are depicted with spheres  with diameters proportional to the ROI size (i.e., the number of voxels).  Notice that the atlas has both cortical and subcortical components.  \textbf{b}: ROI size' distribution and inter-ROI distance' distribution. To give an estimation, as each voxel is about 4 cubic millimeters (see Methods), the ROI average size ($\approx$ 150  voxels) is equivalent to a  3D cube of 21mm edge. Biggest ROI ($\approx$ 600  voxels) corresponds to 3D cubes of 34mm edge. }
\label{fig1}
\end{figure}

Regions of Interest (ROI)  were defined following the Automatic Anatomical Labeling (AAL) atlas \cite{tzourio2002} (see Figs. \ref{fig1}a,  \ref{fig1}b and \ref{fig1}d) which comprises 90 different areas, 45 on each hemisphere (e.g., hippocampus Left, hippocampus Right, amygdala Left, amygdala Right, etc.). Importantly for the study of DOC patients, the AAL atlas includes both cortical and subcortical components (eg., hippocampus, thalamus and amygdala). Per each ROI we  have extracted a mesoscopic (multi-voxel) fMRI time-series  resulting from  averaging over all fMRI time-series of all voxels within a given ROI (Fig. \ref{fig1}b is showing the ROI size distribution among all areas). The MNI coordinates of the centroids in each ROI are used  to calculate the Euclidean distance between each pair of regions (Fig. \ref{fig1}b).

\subsection*{Functional Connectivity matrices}
Correlated areas in the rs-fMRI time series define the Functional Connectivity (FC) matrices. Two methods have been used to  the FC: The Partial Correlation (PC) and the Transfer Entropy (TE). 
\subsubsection*{The Partial Correlation} matrix has dimensions 90$\times$90 (with 90 the  ROIs number) and each element is given by the pairwise PC between any two ROIs. Partial correlation  is a   correlation matrix that removes for a given ROIs pair the effect of the rest of the variables, i.e., removing the correlations  contribution  which are coming from common neighbors interactions. Let   $C$ be a non-singular correlation   matrix, then each element of the PC matrix is given by
\begin{equation} 
 \mathrm{PC}_{ij}=-\frac{P_{ij}}{\sqrt{P_{ii}P_{jj}}}
\label{PC}
\end{equation}
where    $P\equiv C^{-1}$ is the inverse of the correlation matrix (ie., the precision matrix).

Notice that  PC is a symmetrical measure, i.e., $\mathrm{PC}_{ij}=\mathrm{PC}_{ji}$. We also have computed the standard  correlations $C$, and although C is  more noisy than PC, the results we are showing  here for the PC are also valid for the standard correlation.

The PC was computed by using the \textit{partialcorr} method incorporated in MATLAB  (MathWorks Inc., Natick, MA).  The second argument that the function \textit{partialcorr} outputs is  a matrix of p-values for testing the hypothesis of no partial correlation against the alternative that there is a non zero partial correlation.  

PC matrices were calculated for each subject and  grouped into the following categories: inter-hemispheric (between one area on the left and all the other areas at right hemisphere, or vice versa), homologous inter-hemispheric (one area on the left hemisphere and its homologous  area on the right hemisphere, or vice versa), left  intra-hemispheric,  right intra-hemispheric, and total.

\subsubsection*{Transfer Entropy} quantifies  the \textit{directed} interaction  between any two  ROIs. To compute it, let define $i^F$ as the future of the time series in ROI $i$. Similarly, $i^P$ and $j^P$ the pasts of ROIs $i$ and $j$. Then, the TE from $j$ to $i$ is defined as
\begin{equation} 
 \mathrm{TE}_{ji}=H(i^F|i^P)-H(i^F|i^P,j^P)
\label{TE}
\end{equation}
with $H(i^F|i^P)=H(i^F,i^P)-H(i^P)$, the conditional Shannon entropy of  $i^F$ conditioning on $i^P$ (for details, see \cite{cover2006}). Similarly, $H(i^F|i^P,j^P)=H(i^F,i^P,j^P)-H(i^P,j^P)$ is the conditional Shannon entropy of $i^F$ conditioning on $i^P$ and $j^P$. 

The TE is a non-symmetrical measure, i.e., $\mathrm{TE}_{ij}\neq\mathrm{TE}_{ji}$.

The Shannon Entropy (average uncertainty) of the random variable  $X$ is defined as $H(X)=-\sum_x \mathrm{prob}(x) \mathrm{log} \,\,\mathrm{prob}(x) $, where $x$ represents a possible state in variable $X$ \cite{cover2006}. For base 2  logarithm (as we have done here), the information is expressed as  information bits.

To compute probabilities from continuous variables, we did not perform binning; alternatively, we just  rounded each value in the time series to its nearest integer and   computed probabilities (number of time points in a given state  divided by the total time-series length). The conditional entropies have been calculated with the function \textit{condentropy} developed by Hanchuan Peng in C++ and plug-into MATLAB via mex. The  code is available for  download from \cite{pengURL}.

For the past  of the time series it was considered the original  time series. Their future  were  built by shifting the time series in MATLAB with the function \textit{circshift} with a lag  value of $10$ time points. This lag number was previously chosen (and  fixed for all simulations) in order to maximize TE values.

The statistical significance of the TE values was estimated  by shuffling the time series of the target ROI (for the calculation of the TE from $j$ to $i$, hereafter $j$ will be referred as the source and $i$ as the target).  The time series was  shuffled to remove the  temporal information in the target variable. Next, the TE value    is calculated for many repetitions of this shuffling procedure to obtain the distribution of values under the null hypothesis of zero values of TE (i.e., zero uncertainty reduction from source $j$ to target $i$).

TE matrices were calculated for each subject and  grouped into the following categories:  homologous inter-hemispheric (one area on the left hemisphere to its homologous  area on the right hemisphere and vice versa),   left  intra-hemispheric,  right intra-hemispheric,  inter-hemispheric (from one area on the left to all the other areas at right hemisphere, and from one area on the right to all other areas at the left hemisphere)   and total.

\subsubsection*{Summary of brain categories.}
For easy reading  we have adopted the following notation: 

\begin{itemize}
\item PC calculations:

LR: inter- hemispheric (between one area on one hemisphere and all the other areas at the other  hemisphere). As the PC calculation is symmetric (LR is the same than RL) we condensed the inter-hemispheric PC in only LR.

HIH: homologous inter-hemispheric.

LL: left intra-hemispheric.

RR: right  intra-hemispheric.

\item TE calculations:

HLR: homologous inter-hemispheric from left to right (one area on the left hemisphere to its homologous  area on the right hemisphere).

HRL: homologous inter-hemispheric from right to left (one area on the right hemisphere to its homologous area on the left hemisphere).

LL: intra-hemispheric from left to left.

RR: intra-hemispheric from right to right.

LR: inter-hemispheric left-right (from one area on the left to all the other areas at right hemisphere).

RL: inter-hemispheric right-left (from one area on the right to all the other areas at left hemisphere).

\end{itemize}

\subsection*{Statistical analysis}

PC and TE individual matrices were thresholded at a probability value of 0.1 (i.e., $10\%$ confidence); these data were used for  Tables 2 and 3 and all the figures  shown in the paper.  We  also computed PC and TE matrices at different confidence values, $5\%$ and $100\%$ (zero threshold), and the results did not considerably change (cf. Tables S1-S4).

 For comparison of PC and TE values between the different brain categories and groups, a two-ways  ANOVA test was performed, using the function \textit{anovan} from MATLAB   (MathWorks Inc., Natick, MA). For post-hoc analysis, multi-sample t-tests were performed between groups for each brain category using the function \textit{multcompare} from MATLAB which include the Bonferroni correction for multiple comparisons.  To assess possible deviations from the Gaussian distribution in the data, the Kruskal-Wallis non-parametric tests were also performed using the function \textit{kruskalwallis} from MATLAB. The groups comparison results showed very little  differences  across these tests,  cf. Tables 2 and 3, in which the statistically  significant  differences from control were denoted using asterisks at different colors (black for ANOVA and green for Kruskal-Wallis).

\subsection*{A further test for {f}MRI head motion artifacts}

To reject the possibility of head motion artifacts, PC  was re-computed in a matrix which included  the  original 90 ROIs from the AAL atlas plus two motion regressors: the translational modulus and rotational modulus. It is expected, if important correlations were introduced by head motion, that the PC results obtained from this expanded matrix must show significant differences in comparison with the results gathered from the original 90 ROI's. However, this was not the case;  no changes were observed which indicates that the data is free of heat motion artifacts.

\section*{Results}
\subsection*{Partial Linear Correlations (PC)}
\begin{table}
\begin{center}
\caption{\label{Table2}PC average values $\pm$ standard deviation thresholded at $10\%$ confidence (see Methods); \textit{*significantly different from G1; p$<$0.05.} Significant differences are indicated with black asterisks for ANOVA  and green for Kruskal-Wallis tests. LR: inter-hemispheric; HIH:   homologous  inter-hemispheric ; LL: left intra-hemispheric; RR: right  intra-hemispheric. }
\begin{tabular}[t]{|l|l|l|l|l|}
\hline
PC & G1 & G2 & G2a & G2b\\
\hline
LR & 0.11$\pm$0.01 &    0.12$\pm$0.01    & 0.11$\pm$0.01 & 0.12$\pm$0.01 \\
\hline
HIH & 0.40$\pm$0.03 &   0.24$\pm$0.03 * \green{*}   & 0.24$\pm$0.04 * \green{*}  & 0.26$\pm$0.04  *  \green{*}\\
\hline
LL  & 0.13$\pm$0.01 &   0.15$\pm$0.01   \green{*}  & 0.14$\pm$0.09   \green{*}   & 0.15$\pm$0.01  \\
\hline
RR & 0.13$\pm $0.01 &   0.15$\pm$0.01   \green{*}  & 0.14$\pm$0.09  \green{*}   & 0.15$\pm$0.01   \green{*} \\
\hline
Total & 0.12$\pm$0.01 &    0.13$\pm$0.01  *  \green{*}  & 0.13$\pm$0.01   * \green{*}  & 0.13$\pm$0.01  * \green{*}\\
\hline
\end{tabular}
\end{center}
\end{table}

First we looked into the   partial correlation patterns (Table \ref{Table2}). ANOVA between G1 and G2 shows a significant effect of categories (p$<$0.001), and a significant interaction between categories and groups (p$<$0.001). Controls have a significantly smaller PC mean value than patients (p$<$0.001). When looking into  categories,  HIH   PCs are significantly higher than LL, RR, and LR   (p$<$0.001). In addition, LL and RR values are significantly higher than LR (p$<$0.001). 
To further inspect the interaction, we performed post-hoc multiple comparison tests between groups for the different categories. HIH PCs are significantly higher in G1  (p$<$0.001). 
The Kruskal-Wallis test gave the same results, with the addition of being LL and RR PCs significantly higher in G2 compared with G1(p$<$0.005).

\begin{figure}[h]
\centering{
\includegraphics[width=.75\textwidth]{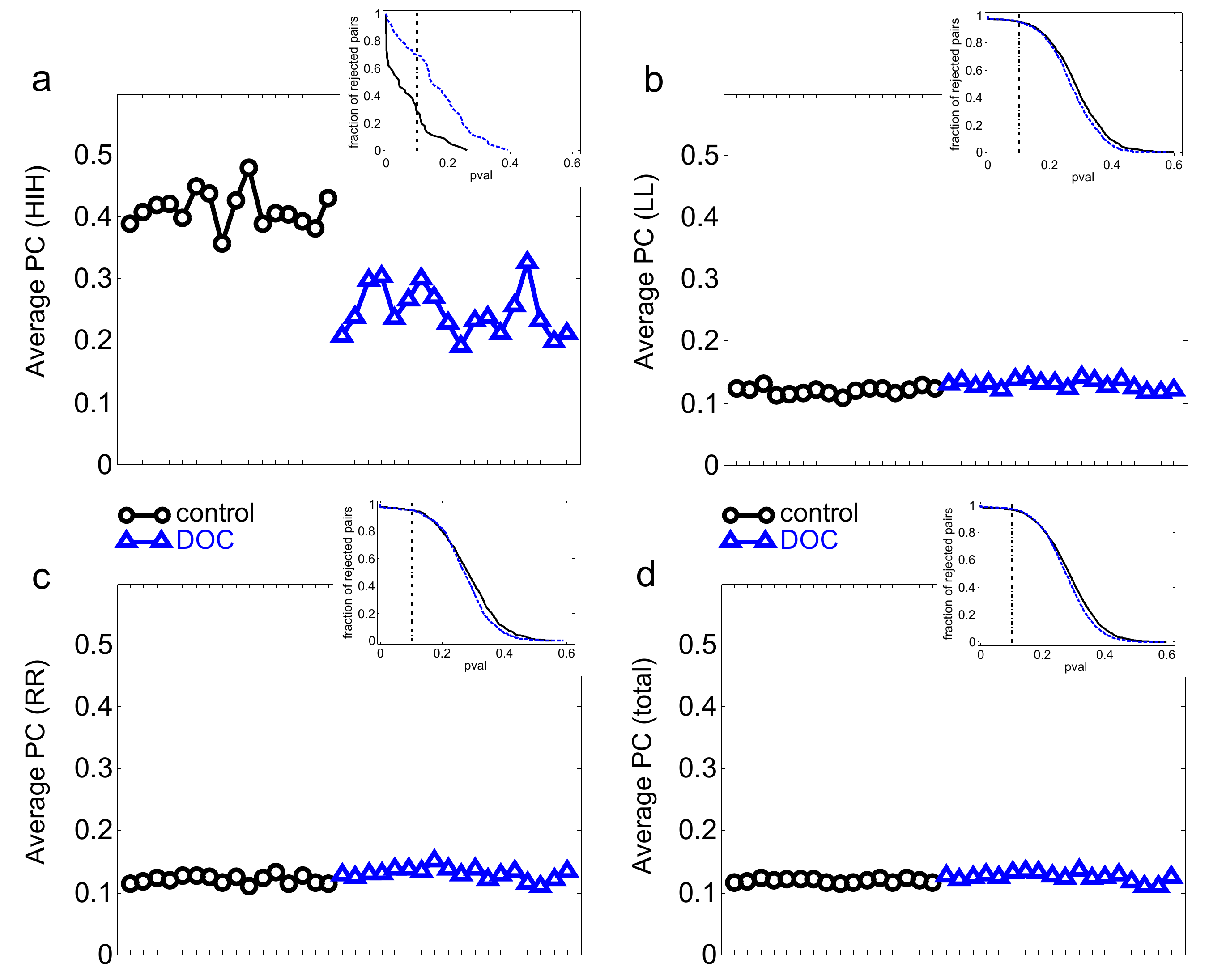}
}
\caption{\textbf{Average PC values per subject}. \textbf{a}: HIH (homologue  inter-hemispheric   areas); \textbf{b}: LL (left intra-hemispheric); \textbf{c}: RR (right  intra-hemispheric); \textbf{d}: total. Insets depict the fraction of rejected pairs of areas for a given probability level. PC values  were thresholded at a probability value of 0.1 (dashed lines in the insets) . Black circle: G1 (control); blue triangles : G2 (DOC). Observe the huge differences between G1 and G2 for HIH compared to LL and RR. For detailed values, see Table \ref{Table2}.}
\label{Figure2}
\end{figure}

The comparison between G1 and G2a gives the same results. However, when comparing G1 and G2b, the effect of group still holds but is smaller than that between G1 and G2a (p$=$0.002). The effect of categories  is the same as in G1 vs. G2 comparison, and there is a significant interaction effect (p$<$0.001). Post-hoc tests show that HIH PCs are significantly smaller in G2b with respect to G1 (p$<$0.001). Finally, the comparison including all brain categories (total) was significant between G1 vs G2 and G1 vs G2a (P = 0.018 and 0.044 respectively).  The same significant differences were conserved with the Kruskal-Wallis test. Results can be seen in Table \ref{Table2} and Fig. \ref{Figure2}.

In summary, the partial linear correlations approach allows to expose a differential functional connectivity in a healthy conscious brain in comparison with a DOC state  and a recent recovery  from it. A reduced inter-hemispheric connectivity is  evident in DOC patients. 

\subsection*{Transfer Entropy}
We then examined the uncertainty reduction (information) transferred between ROIs pairs by computing the transfer entropy  (TE). ANOVA on TE values for G1 and G2 shows a significant effect of group (p$=$0.0025) and categories  (p$<$0.001). Particularly there were significant differences between HLR and HRL TEs and the TE values for the other categories. In the case of HLR, TEs are significantly lower than LL, RR and inter-hemispheric (LR and RL) TEs (p$<$0.005), whereas HRL TEs are significantly lower than RR and RL TEs (p$<$0.025). There is no interaction effect between group and brain category. The post-hoc analysis showed that LL, LR and the total TE values differ between controls and patients. 
 
However, when performing the ANOVA for G1 and G2a there is a significant effect of group (p$<$0.001) and categories  (p$<$0.001). 
Post-hoc tests show that LL and RR TEs are significantly higher in G1 than in G2a (p$<$0.05). In addition, TE for LR is also significantly higher in G1 (p$=$0.001).
 
When comparing G1 and G2b, there was significant effect of group (p$=$0.042) and categories  (p$<$0.001). Additionally, when looking into the main effect of brain sections, HLR and HRL TE values were significantly smaller than LR and RL (p$<$0.05). However the multiple-compare test did not revealed any significant difference.
The Kruskal-Wallis test gave the same general results but in this case adding significant differences between G2 and G1 in the same regions were  we previously found only for G2a.

\begin{figure}[h]
\centering{
\includegraphics[width=.75\textwidth]{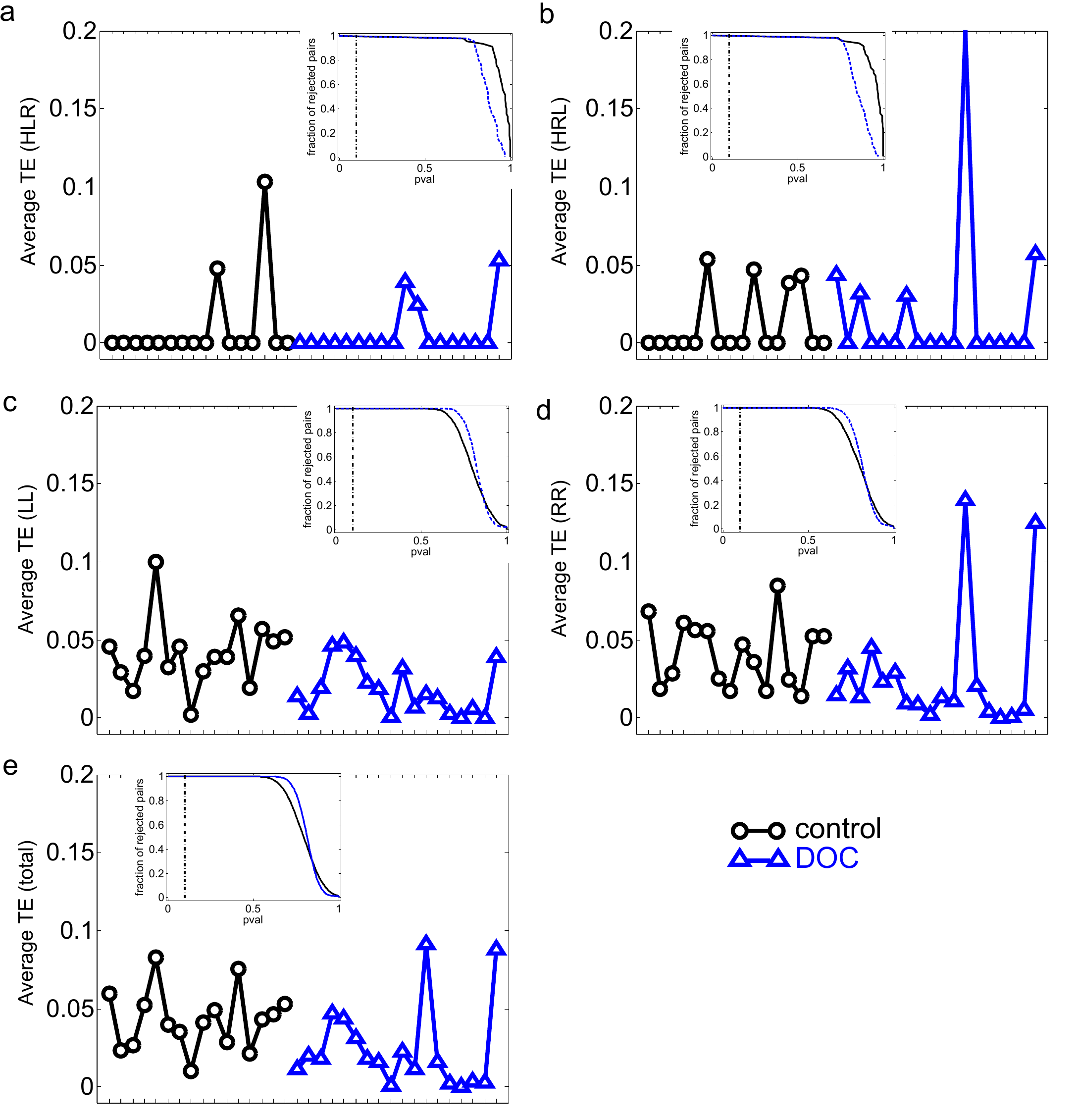}
}
\caption{\textbf{Average TE values per subject}. \textbf{a}: HLR (homologue   left-right inter-hemispheric   areas);  \textbf{b}: HRL (homologue right-left  inter-hemispheric   areas);  \textbf{c}: LL (left intra-hemispheric); \textbf{d}: RR (right  intra-hemispheric); \textbf{e}: total.  Insets depict the fraction of rejected pairs of areas for a given probability level.  TE values  were thresholded at a probability value of 0.1 (dashed lines in the insets). Black circles: G1 (control); blue triangles : G2 (DOC). 
 }
\label{Figure3}
\end{figure}

The results can be seen in Table \ref{Table3} and Fig. \ref{Figure3}. If two time series are highly correlated, their TE is close to zero in both directions; if they are not correlated but one influences  the other's behavior, TE is high in that direction and very  low in the opposite direction. In our results, the significant smaller TE  between homologue areas with respect to the other TE values is consistent to the fact that they are highly correlated (cf.  results in 3.1).

The differences found within hemispheres between the groups parallelize the increased intra-hemispheric correlations in G2 and G2a. When looking at G2b group, their averages are also biased by one patient that presented extremely high TE values (corresponding to the last case in the x-axis).

\begin{table}
\begin{center}
\caption{\label{Table3}  TE average values $\pm$ standard deviation. \textit{*significantly different from G1; p$<$0.05.} Significant differences are indicated with black asterisks for ANOVA  and green for Kruskal-Wallis tests.  HLR: homologous inter-hemispheric from left to right;  HRL: homologous inter-hemispheric from right to left;  LL: left intra-hemispheric; RR: right  intra-hemispheric; LR: inter-hemispheric left to right; RL: inter-hemispheric right to left. }
\begin{tabular}[t]{|l|l|l|l|l|}
\hline
TE & G1 & G2 & G2a & G2b\\
\hline
HLR& 0.009$\pm$0.027               & 0.006$\pm$0.015                   & 0.004$\pm$0.011                        & 0.017$\pm$0.030 \\
\hline
HRL  & 0.011$\pm$0.020                & 0.020$\pm$0.049                     & 0.020$\pm$0.053                           & 0.019$\pm$ 0.032\\
\hline
LL & 0.040$\pm$0.021                  & 0.017$\pm$0.016  *\green{*}                 & 0.013$\pm$0.013 *  \green{*}                      & 0.040$\pm$0.003\\
\hline
RR & 0.039$\pm$0.020                  & 0.027$\pm$0.039          \green{*}         & 0.019$\pm$0.033 *  \green{*}                       & 0.065$\pm$0.050\\
\hline
LR & 0.043$\pm$0.024                  & 0.021$\pm$0.021  *\green{*}               & 0.016$\pm$0.017 *  \green{*}                     & 0.047$\pm$0.017\\
\hline
RL & 0.043$\pm$0.021                    & 0.031$\pm$0.045        \green{*}            & 0.024$\pm$0.042            *  \green{*}           & 0.066$\pm$0.049\\
\hline
Total & 0.041$\pm$0.018                  & 0.024$\pm$0.026               *  \green{*}        & 0.018$\pm$0.022 *  \green{*}               & 0.054$\pm$0.028\\
\hline
\end{tabular}
\end{center}
\end{table}

In summary, TE analysis exposes alterations in the FC exhibited by DOC patients. In particular, TE  within hemispheres and between hemispheres is smaller, although no difference was found when looking at homologue areas.   In contrast to the results obtained in the PC analysis, the differences found uphold irrespective of the Euclidean distance separating ROIs pairs, although when considering LL TE, a slight decrease in the statistical p value can be observed.

\subsection*{Between-homologue inter-hemispheric  PC and left intra-hemispheric TE}

\begin{figure}[h]
\centering{
\includegraphics[width=.75\textwidth]{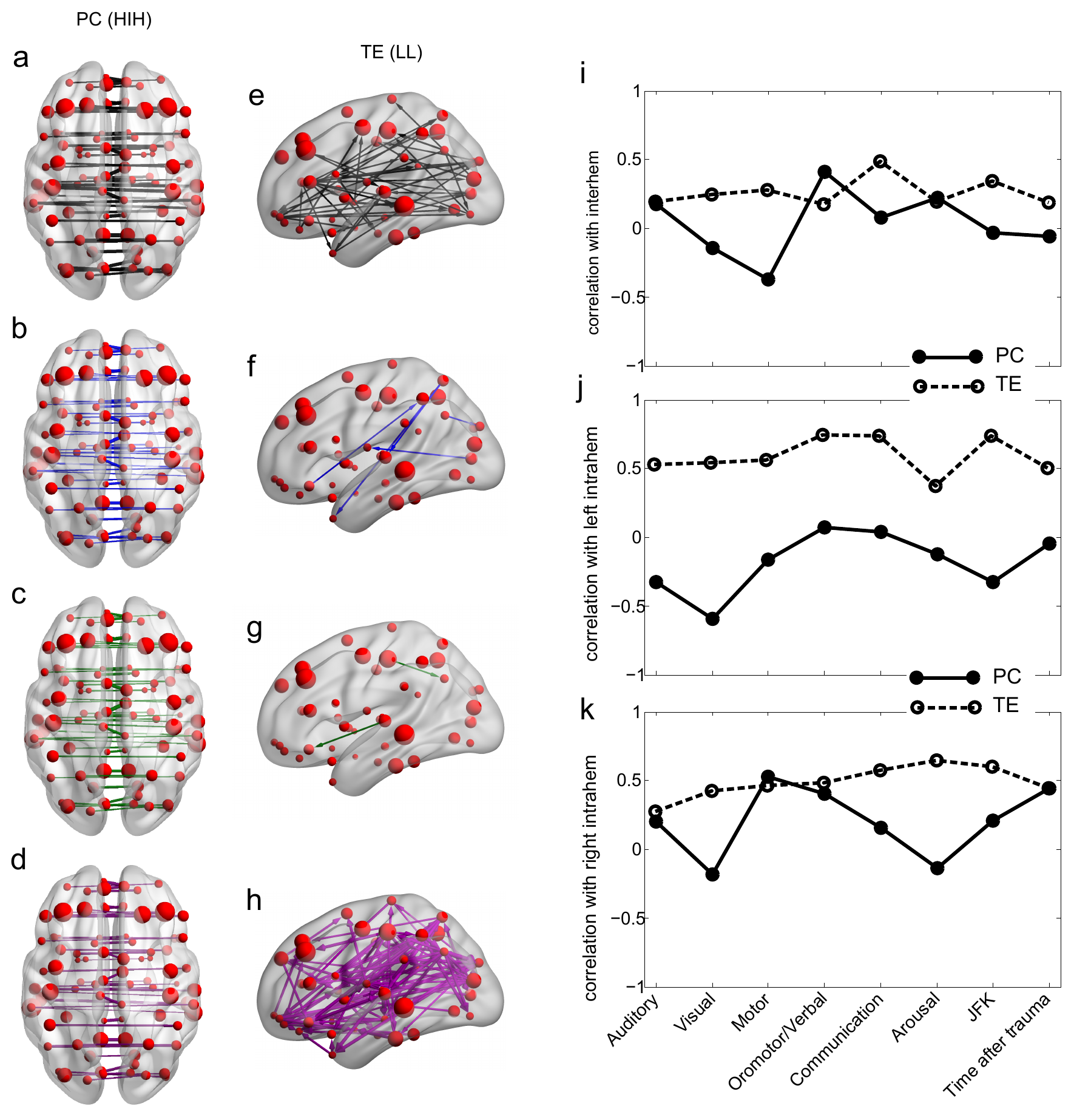}
}
\caption{\textbf{Inter-hemispheric PC and left intra-hemispheric TE.} \textbf{a-h}: PC and TE values for all the 4 different groups. The thickness of links and arrows are proportional to the PC and TE values; the thickness normalization factor is common among all the 4 groups. \textbf{a,e}: group G1, black, \textbf{b,f}: group G2, blue, \textbf{c,g}:  group G2a, green,   \textbf{d,h}:   group G2b, magenta.   \textbf{a-d}: Visualization of the PC values HIH (homologue inter-hemispheric pairs).   \textbf{e-h}: TE in  LL (left intra-hemispheric pairs). For clarity in the visualization, links have been thresholded and  only TE values bigger than TE=0.2 are depicted.  \textbf{i-k}: Correlation between PC (solid line) and TE (dashed) with the CRS-R scores at   the   different functional scales: Auditory, Visual,Motor, Oromotor/Verbal,  Communication, Arousal and the total sum over all the function scales (JFK) as well as with the acquisition time after trauma.   The correlation has been calculated  over pairs which are \textbf{i}:  inter-hemispheric (HIH  for PC and (HLR+HRL)/2 for TE, \textbf{j}: left intra-hemispheric (LL) and    \textbf{k}: right  intra-hemispheric (RR).   
 }
\label{Figure4}
\end{figure}

The results show that for all analyzed areas the best two discriminators  are the  between-homologue inter-hemispheric (HIH) PC (Figs. \ref{Figure4}a-d) and the  left intra-hemispheric (LL) TE  (Figs. \ref{Figure4}e-h). Here, colors denote group differences: black (G1), blue (G2), green (G2a) and magenta (G2b). For both PC and TE the thickness of links and arrows is proportional to the PC and TE values.

For PC there is a manifest anatomical disparity in the correlations pattern: it can be observed that homologue areas that are closer to each other show stronger correlations than farther ones (i.e. thicker connections at shorter distances in comparison with thinner connections at longer distances). To disentangle the behavior of the neural correlations regarding to a spatial factor, we look at the Euclidean distances  between the centroids of homologue areas. For G1 the areas close to each other presented a high correlation, and beyond a threshold distance of 20 mm, correlations decreased, although the values remained high. Interestingly, the same behavior was  found in G2. However, the correlation values there were  shifted down, with lower mean value for areas closer than 20 mm, and decreasing for increasing distances.  Thus,  for ROIs areas distance-separated smaller than 20mm, differences between G1 and G2 were smaller compared to areas separated at long distances, distance separation $<$ 20mm pval=$10^{-6}$,  distance $>$40mm pval=$10^{-14}$. 
When inspecting G2a and G2b subgroups, there were no observable differences for anatomically closer areas, whilst  it could  be detected a higher correlation of some of the anatomically further areas for G2b.

 Regarding to the TE, not only the mean values of TE in LL areas were different between groups (Table \ref{Table3}), but the number of significant values of TE, i.e.,  the number of arrows plotted in Figs. \ref{Figure4}e-h varies across different groups. This number 
was more  than 9 times bigger in G1 compared  with  G2 (G1   \# links=47,   Fig. \ref{Figure4}e     ;     G2     \# links=5,   Fig. \ref{Figure4}f).  When comparing with  group G2b, this number doubled the one in group G1  (\# links=99, Fig. \ref{Figure4}h), possibly indicating a   "transient" brain state in the pattern of  information flows in group G2b in comparison with control.

\subsection*{Correlation between fMRI measures and   CRS-R scores}

We then asked if the two fMRI measures, between-homologue inter-hemispheric  PC   and    left intra-hemispheric TE were correlated with the neurological and behavioural scale given by the CRS-S. This is represented in Figs. \ref{Figure4}i-k. For homologue inter-hemispheric pairs we found that TE gave  the biggest correlation  with the corresponding  value in the communication function scale.  For left intra-hemispheric pairs, TE had  0.73 correlations with oromotor/verbal function scale, 0.73 with the communication function scale and 0.73 with the total CRS-R (marked as "JFK" in Figs. \ref{Figure4}i-k).

\section*{Discussion} 

In this study we have  investigated whether  the functional   connectivity   is altered as a consequence of consciousness disturbances. We have applied the Partial Correlation   and the Transfer Entropy   approaches to analyze the FC  from resting-state fMRI data. We have compared two groups,  healthy subjects   and  Disorder of Consciousness   patients.  The analysis was done over the 90 anatomical brain areas, defining  regions of interest   from the AAL atlas. We have grouped the different pairs of ROIs in inter-hemispheric homologue regions, inter-hemispheric , left intra-hemispheric, right intra-hemispheric  and total (all regions). We have found two particular markers that account for the large-scale disturbance of patients brain function: the PC calculated over homologue inter-hemispheric (HIH) regions and the TE calculated over the left intra-hemispheric (LL)  ROIs. 

The PC in HIH regions was found to be notably larger for control compared to DOC patients. This results holds also when comparing G1 with the recovered G2b group. The same comparison but done over the total average of the 90 regions did not shown significant differences. Thus, one relevant  result of our analysis 
is the finding that only by the calculation of the PC in the proposed grouping of brain regions, it was possible to detect a significant marker for the patients disturbance, results that is hidden when we looked at the PC of the total  AAL brain regions. 

In the case of TE, the total score did not show any significant difference either, but the brain subdivision revealed that the  intra-hemispheric influences were different in control respect DOC.  This happened for both LL and RR, although the TE in LL discriminated  better than in  RR.  This is a very novel finding whose origin is still unclear and deserves further investigation.

\subsection*{Methodological issues}
The PC is a straightforward  measure  able to eliminate for each specific ROIs pair, the contribution to the correlations coming from common neighbors, preserving \textit{effective} correlations between two time series. Unlike the PC which is  a symmetrical measure,  the TE quantifies interaction between ROIs   in a directed form, i.e., region A influences to region B but the opposite is not necessary true. In concrete, TE quantifies information bits (uncertainty reduction)  flowing  from one ROI to the future of the other.   For the case of Gaussian data, the information bits measured by the TE coincide with the Granger causality measured from   time series \cite{barnett2009};  however for non Gaussian data, TE and causality might result in different measures.  

TE emerges as a very suitable measure for the study of temporal causality in brain fMRI activity in parallel to the advantage of an accurate spatial resolution. TE assessment in a population of patients with disorder of consciousness provides the opportunity of gaining insight into brain mechanisms of information processing and the finding of possible predictors of coma outcome.

Regarding to the calculation of TE, it is well-known that the computation  of the entropies with small data sets introduces some a bias \cite{panzeri1996,paninski2003,bonachela2008}. Because we are performing groups comparison with the same data size in each group (i.e., the time series in each subject have the same data points), such a bias will be the same  in the two groups, thus not affecting the validity of    the groups comparison. 
Nevertheless, as far as we understand  there is not any  reported study  analyzing either information reduction (i.e. TE) or causality  in fMRI data from DOC patients.

\subsection*{Inter-relation between PC and TE in DOC patients}
To exhibit high correlations is different from having high  TE between two time series.   This can be   clearly understood by a counter-example; two fully correlated time series have zero TE  as to compute  the uncertainty reduction in the future of $i$,  conditioning on the two pasts $i$ and $j$  is not adding any further information to the situation of solely adding the past of $i$, i.e. the two terms in the right-hand side in Eq. (\ref{TE}) are equal.
As a consequence of this, the observation of having  high  PC  for   HIH pairs   in healthy subjects implies   to have  high isolation of the information  within hemispheres; thus, the TE values in both   LL and RR are significantly higher than the corresponding values in HLR and HRL. 

Interestingly, we found that while PC is reduced in DOC patients between inter-hemispheric homologue areas, TE shows an altered pattern at the level of general inter-hemispheric interactions. In the control group we observe that despite the coherence is high between homologue areas, their TE is low. Conversely, while PC between hemispheres is low, LR and RL TE are high. The DOC patients show the same trend, although the LR and RL TE is significantly lower than in controls. This supports the notion that consciousness arises from long-range modulation of neural activity. A disruption in long-range communication could affect mechanisms such as increase of stimulus' salience, facilitation of propagation across sparsely connected networks, and selective routing \cite{ganzetti2013}, mechanisms that are related to conscious processing \cite{gaillard2009}.

\subsection*{rs-fMRI inter-hemispheric correlations and gamma rhythms}
Recently it has been shown that the inter-hemispheric correlations in the rs-fMRI dynamics   correlate with  the inter hemispheric coherence exhibited by  electrophysiological recordings in human sensory cortex \cite{nir2008}, mainly  with the slow modulation of the gamma rhythms in Local Field Potentials.  Other studies have also found such modulation in high-level cognition tasks \cite{vidal2012}. Thus, one could conjecture that at the functional level, a breakdown in the inter-hemispheric rs-fMRI correlations in DOC patients could be  an indication of a similar deficit in  the  gamma power coherence.   One possibility is that low-frequency oscillatory activity is related to an underlying neuronal mechanism allowing for maintenance and consolidation of neural events across wide sections of the brain, and for the handling of incoming stimuli \cite{depasquale2010, depasquale2012}. Although increasing evidence points toward a property of the brain relevant for conscious processing, Vidal et al. \cite{vidal2012} point out that gamma-amplitude correlation would also be reflecting the parallel organization of the brain, where neural networks interact for purposeful processing of information.

\subsection*{Comparison with previous results}

As fas as we know, a single  study have reported  that DOC patients in comparison with healthy subjects manifest a strong reduction in the inter-hemispheric correlations in the rs-fMRI time series \cite{ovadia2012}.  The authors in  \cite{ovadia2012} did not use any atlas to compute inter-hemispheric correlations; instead they investigated  specific areas such as pre- and post-central gyrus and  the intra-parietal sulcus. Among other reasons, the authors selected  those areas for being well   separated each from the other (arguing the existence of less noise in the signal).  This is consistent with our finding that DOC patients kept more similar correlations to control for  ROIs separation below 20 mm. In addition to this, our study adds the novelty of having analyzed the FC  obtained by the TE.

\subsection*{TE density to measure consciousness alteration}
We have shown in Figs. \ref{Figure4} e-h how the number of TE connections can account not only  for  the differences between control (G1) and DOC (G2)   but for the   transitory  brain state in the group G2b: the   patients that awaked and became  fully conscious   at the second fMRI acquisition. Thus, we have found that the number of TE connections were 47 (G1), 5 (G2) and 99 (G2b). In a similar spirit,  Seth and colleagues  \cite{seth2006} defined the causal density for measuring consciousness  in brain states  as the number of Granger-causality connections flowing in and out per each specific area. Interestingly, a similar behavior has been reported during recovery from anesthesia, where an increment in functional connectivity above the normal wakeful baseline is found \cite{hudetz2012}.

\subsection*{DOC impairment at specific brain areas}

The aim of this analysis   is not to work   at the level of an  individual DOC patient  but to search for  rs-fMRI markers that can account for groups differences in DOC patients. We have not studied yet any  measure that can account for DOC impairment at specific brain areas.  To this end, one could study in principle the FC graphs obtained by either PC or TE using complex networks analysis, or any other kind of graph exploration methods. In a much simpler spirit (just to illustrate that this approach is plausible), we have chosen to plot the PC values   per area comparing group G2 versus G1.  This is illustrated in the Suppl. Fig. S1. The decorrelation index per area is plotted, (corrG1-corrG2)/corrG1. Colored  in blue,   the five biggest decorrelation indices correspond to the following areas: Fusiform, Insula, Parietal Superior, Precentral and Temporal Superior, revealing that those areas had the major DOC impairment. Conversely the areas with less DOC impairment (colored in red) were the Cingulum Anterior, Cingulum Middle, Frontal Superior Orbital, Superior Motor Area and Temporal Inferior.

\subsection*{Limitations of the study}
One of the important limitation of studying DOC patients is the great amount of involuntary movements they exhibit, leading to potential  artifacts in the fMRI acquisition. Techniques to overcome this issue include  affine transformations   to the time series  creating a head-motion parameter matrix  which can be used to regress out and  remove the spurious variances they introduce \cite{fox2005}. Although these methods can correct  signals from movements spanning the dimensions of up to 3 to 4 voxels,  recent work \cite{power2012} suggest that no technique could remove completely the effects of these artifacts over the FC.  Thus especial care is necessary to tackle these problems and, eventually, discard the entire scan.

\subsection*{Future directions}
In this study  PC and TE  measures  were used to assess for the assessment of functional  connectivity in unconscious patients. In particular we characterized their disruptions at an anatomical level, in the basis of distances between homotopic areas. Other questions that can be explored, include  the integrity of FC between the areas that constitute hubs in the   brain network, between areas with high \textit{rich-clubness} \cite{vandenheuvel2011}, or between associative vs sensory areas . 

\subsection*{Funding}
J.M.C. is  supported  by Ikerbasque: The Basque Foundation for Science.  J.M.C.  acknowledges financial support from Junta de Andalucia, grant P09-FQM-4682. V. M-M, M.V, and D.R.C. are partially supported by CONICET (National Council of Scientific and Technological Research) of  Argentina.  Additional support was provided by the Dept. of Neurology, and Dept. of Teaching and Research of FLENI, Buenos Aires, Argentina.

\clearpage

\clearpage

\clearpage

\clearpage

\setcounter{figure}{0}

\renewcommand{\thefigure}{S\arabic{figure}}

\clearpage
\begin{figure}[h]
\centering{
\includegraphics[width=.75\textwidth]{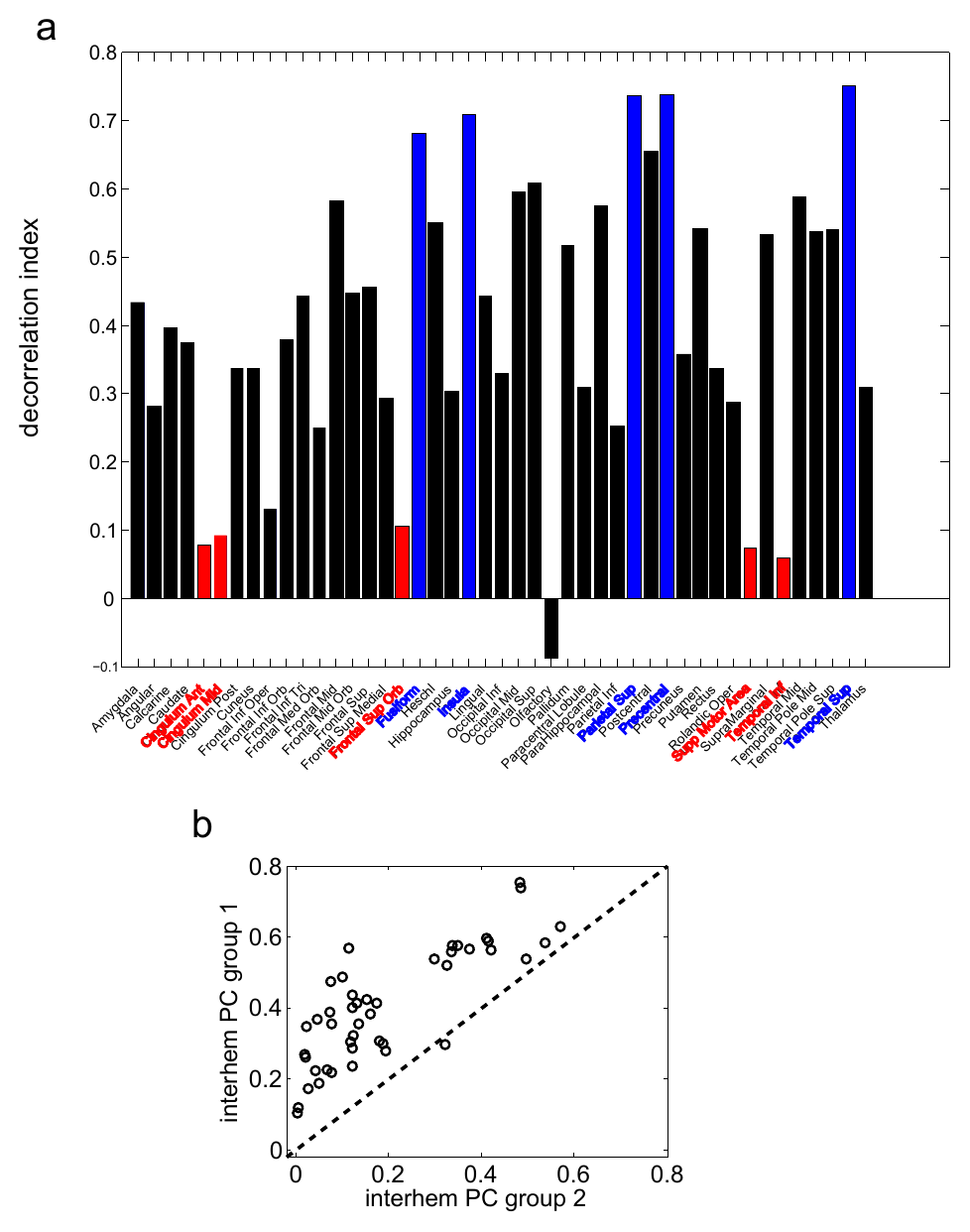}
}
\caption{\textbf{DOC impairment evaluated at specific brain areas} \textbf{a}: Decorrelation indices  defined as (corrG1-corrG2)/corrG1 computed for each of the different 45 brain areas. In blue, top-five  values of  decorrelation index; in red, bottom five (positive) values.   \textbf{b}:  Scatter of between-homologue inter-hemispheric correlations G1 vs G2, each point represents one of the 45 brain areas.
 }
\label{FigS1}
\end{figure}

\clearpage
\setcounter{table}{0}

\renewcommand{\thetable}{S\arabic{table}}

\begin{table}
\begin{center}
\caption{\label{TableS1} PC average values $\pm$ standard deviation thresholded at $5\%$ confidence. \textit{*significantly different from G1; p$<$0.05.} Significant differences are indicated with black asterisks for ANOVA  and green for Kruskal-Wallis tests.  LR: inter-hemispheric; HIH: between-homologue inter-hemispheric ; LL: left intra-hemispheric; RR: right  intra-hemispheric. }
\begin{tabular}[t]{|l|l|l|l|l|}
\hline
PC & G1 & G2 & G2a & G2b\\
\hline
LR & 0.10$\pm$0.004 &    0.11$\pm$0.008     & 0.10$\pm$0.009 & 0.11$\pm$0.002   \green{*}\\
\hline
HIH & 0.41$\pm$0.031 &   0.24$\pm$0.041 * \green{*}   & 0.24$\pm$0.045 *  \green{*}  & 0.25$\pm$0.053  *  \green{*}\\
\hline
LL  & 0.11$\pm$0.006 &   0.12$\pm$0.008    \green{*}  & 0.12$\pm$0.008   \green{*}   & 0.12$\pm$0.009    \green{*}\\
\hline
RR & 0.11$\pm $0.007 &   0.12$\pm$0.010     \green{*}  & 0.12$\pm$0.011   \green{*}   & 0.12$\pm$0.003   \green{*} \\
\hline
Total & 0.11$\pm$0.004 &    0.11$\pm$0.008  *  \green{*}  & 0.11$\pm$0.008   \green{*}  & 0.12$\pm$0.003  *  \green{*}\\
\hline
\end{tabular}
\end{center}
\end{table}

\clearpage

\begin{table}
\begin{center}
\caption{\label{TableS2} PC average values $\pm$ standard deviation  thresholded at 100\% confidence (i.e., zero threshold);  \textit{*significantly different from G1; p$<$0.05.} Significant differences are indicated with black asterisks for ANOVA  and green for Kruskal-Wallis tests. LR: inter-hemispheric; HIH: between-homologue inter-hemispheric ; LL: left intra-hemispheric; RR: right  intra-hemispheric. }
\begin{tabular}[t]{|l|l|l|l|l|}
\hline
PC & G1 & G2 & G2a & G2b\\
\hline
LR &   0.16$\pm$0.003 &    0.16$\pm$0.006    & 0.16$\pm$0.007 & 0.17$\pm$0.002 * \green{*}\\
\hline
HIH & 0.42$\pm$0.028 &   0.28$\pm$0.036 *  \green{*}   & 0.28$\pm$0.034 *  \green{*}  & 0.29$\pm$0.044  * \green{*}\\
\hline
LL  & 0.16$\pm$0.005 &   0.17$\pm$0.007  *  \green{*}  & 0.17$\pm$0.006  *  \green{*}   & 0.17$\pm$0.008  * \green{*}\\
\hline
RR & 0.16$\pm $0.005 &   0.17$\pm$0.008   *  \green{*}  & 0.17$\pm$0.008  *  \green{*}   & 0.17$\pm$0.004  *  \green{*} \\
\hline
Total & 0.16$\pm$0.003 &    0.17$\pm$0.006  *  \green{*}  & 0.17$\pm$0.006   * \green{*}  & 0.17$\pm$0.003  * \green{*}\\
\hline
\end{tabular}
\end{center}
\end{table}

\clearpage

 \begin{table}

\begin{center}

\caption{\label{TableS3}  TE average values $\pm$ standard deviation thresholded at $5\%$ confidence.   \textit{*significantly different from G1; p$<$0.05.} Significant differences are indicated with black asterisks for ANOVA  and green for Kruskal-Wallis tests. HLR: homologous inter-hemispheric from left to right;  HRL: homologous inter-hemispheric from right to left;  LL: left intra-hemispheric; RR: right  intra-hemispheric; LR: inter-hemispheric left to right; RL: inter-hemispheric right to left. }

\begin{tabular}[t]{|l|l|l|l|l|}

\hline

TE & G1 & G2 & G2a & G2b\\

\hline

HLR& 0.003$\pm$0.014               &  N/A             &   N/A              &  N/A \\

\hline

HRL  & 0.006$\pm$0.017               & 0.008$\pm$0.024                     & 0.004$\pm$0.013                          &  N/A \\

\hline

LL & 0.020$\pm$0.015                 & 0.007$\pm$0.007  * \green{*}                 & 0.004$\pm$0.004 *  \green{*}                      & 0.016$\pm$0.003\\

\hline

RR & 0.019$\pm$0.013                  & 0.013$\pm$0.021          \green{*}         & 0.004$\pm$0.004 *  \green{*}                       & 0.032$\pm$0.031\\

\hline

LR & 0.020$\pm$0.015                 & 0.008$\pm$0.010  * \green{*}               & 0.004$\pm$0.004 *  \green{*}                     & 0.019$\pm$0.011\\

\hline

RL & 0.020$\pm$0.014                   & 0.014$\pm$0.025        \green{*}            & 0.005$\pm$0.006           *  \green{*}           & 0.026$\pm$0.021\\

\hline

Total & 0.020$\pm$0.012                & 0.010$\pm$0.013               *  \green{*}        & 0.004$\pm$0.003 *  \green{*}               & 0.023$\pm$0.015\\

\hline

\end{tabular}

\end{center}

\end{table}

\clearpage

\begin{table}
\begin{center}
\caption{\label{TableS4}  TE average values $\pm$ standard deviation thresholded at 100\% confidence (i.e., zero threshold);   \textit{*significantly different from G1; p$<$0.05.} Significant differences are indicated with black asterisks for ANOVA  and green for Kruskal-Wallis tests. HLR: homologous inter-hemispheric from left to right;  HRL: homologous inter-hemispheric from right to left;  LL: left intra-hemispheric; RR: right  intra-hemispheric; LR: inter-hemispheric left to right; RL: inter-hemispheric right to left. }
\begin{tabular}[t]{|l|l|l|l|l|}
\hline
TE & G1 & G2 & G2a & G2b\\
\hline
HLR& 1.808 $\pm$0.154               & 1.392$\pm$0.397     * \green{*}              & 1.266 $\pm$0.400     * \green{*}                       & 1.729$\pm$0.302  \\
\hline
HRL  & 1.793$\pm$0.163                & 1.444$\pm$0.475        * \green{*}                 & 1.288$\pm$0.463          *  \green{*}                  & 1.783$\pm$  0.378\\
\hline
LL & 1.933$\pm$0.182              & 1.439$\pm$0.439 * \green{*}                 &1.297$\pm$0.430 *  \green{*}                      & 1.844$\pm$0.374\\
\hline
RR & 1.923$\pm$0.187                  & 1.471$\pm$0.492     *  \green{*}              & 1.311$\pm$0.483 * \green{*}                       &1.859$\pm$0.400\\
\hline
LR & 1.938$\pm$0.184                  & 1.433$\pm$0.431 *  \green{*}               & 1.295$\pm$0.427 *  \green{*}                     & 1.832$\pm$0.362\\
\hline
RL & 1.924$\pm$0.190                  &1.486$\pm$0.506          * \green{*}             &1.318$\pm$0.490            *  \green{*}           & 1.886$\pm$0.425\\
\hline
Total & 1.929$\pm$0.183                 & 1.457$\pm$0.464            *  \green{*}        & 1.305$\pm$0.454 * \green{*}               & 1.855$\pm$0.389\\
\hline
\end{tabular}
\end{center}
\end{table}

\end{document}